\documentclass[pre,aps,superscriptaddress,showpacs,twocolumn]{revtex4}
\usepackage{graphicx}
\usepackage{amssymb,amsmath}
\usepackage{color}
\usepackage[T1]{fontenc}

\begin{document}

\title{Influence of fine particles on the stability of a humid granular pile}

\author{Xixi Huang}
\affiliation{Laboratoire de Tribologie et Dynamique des Syst\`emes;
Universit\'e de Lyon; \'Ecole Centrale de Lyon;
CNRS UMR 5513; 36, av. Guy de Collongue, F-69134 \'Ecully, France},
\affiliation{Institut Lumi\`ere Mati\`ere;
Universit\'e de Lyon; Universit\'e Claude Bernard Lyon 1;
CNRS UMR 5306; Domaine scientifique de la Doua, F-69622 Villeurbanne, France}
\author{Sandrine Bec}
\affiliation{Laboratoire de Tribologie et Dynamique des Syst\`emes;
Universit\'e de Lyon; \'Ecole Centrale de Lyon;
CNRS UMR 5513; 36, av. Guy de Collongue, F-69134 \'Ecully, France},
\author{Jean Colombani}
\email{jean.colombani@univ-lyon1.fr}
\affiliation{Institut Lumi\`ere Mati\`ere;
Universit\'e de Lyon; Universit\'e Claude Bernard Lyon 1;
CNRS UMR 5306; Domaine scientifique de la Doua, F-69622 Villeurbanne, France}

\begin{abstract}
We have investigated the influence of a small amount of fine particles on the stability of a granular heap by rotating drum experiments.
The fine particles are shown to have a strong and ambivalent influence.
For low fine particles contents, the heap destabilizes by avalanches, and the fine particles tend to fluidize the heap.
On the contrary, for high fine particles contents, they increase the cohesion of the heap, which destabilizes through stick-slip at the drum wall.
We interpret and model the fluidification in the avalanching regime, which we show is independent of humidity, by granular lubrication considerations, whereas the stick-slip behavior, highly dependent on humidity, is understood by a solid friction model.
\end{abstract}

\maketitle

\section{Introduction}

The stability of humid granular media has been studied extensively during the last years and the influence of the liquid content \cite{Hornbaker,Tegzes,Scheel}, liquid surface tension \cite{Nowak}, grain roughness \cite{Halsey,Mason}, or pile age \cite{Bocquet} have found explanations.
In these studies, the working granular pile is a monodisperse set of spherical grains.
Besides, the study of bidisperse granular media is devoted to the investigation of segregation and mixing \cite{Geromichalos}.
But monodisperse granular media are hardly ever really monodisperse and contain dust due to the wear of the grains \cite{Restagno}.
This is always true not only in sandcastle building on the beach, but also in industrial use of granular media (material grinding, cereal storage, pharmaceutical manufacturing, \dots).

Adding a small controlled amount of micron-size particles to a granular pile, we show by rotating drum experiments that these fine grains have a dramatic effect on the stability of the pile.
Furthermore, this influence is ambivalent: the fine particles fluidify the heap at low content, and solidify it at higher content.
We propose a microscopic mechanism and a theoretical interpretation of this non-monotonic behavior, based on capillarity, jamming, lubrication, and friction considerations.

\section{Materials and method}

The experimental setup consists of a stainless steel rotating drum of 10 cm radius and width, with glass windows on the sides.
The windows have a hole in the center to allow air exchange with the outside atmosphere.
The drum is partially filled with soda-lime glass beads of (250$\pm$100) $\mu$m diameter mixed with small beads of (5$\pm$4) $\mu$m diameter.
The pile is illuminated from the front, observed with a camera, and its surface is detected by an automatic procedure.
Its angle $\theta$ with the horizontal is measured with a 0.5$^\circ$ accuracy.
The drum is placed inside a closed vessel where a constant relative humidity $H$ is maintained by the presence of a wide beaker of saturated aqueous inorganic salt solution.
Each salt fixes the relative humidity in the atmosphere at a given value (6\% with LiBr, 23\% with KCH$_3$CO$_2$, 43\% with K$_2$CO$_3$, 58\% with NaBr).
The values of $H$ are measured with an accuracy of 1.5\% by an hygrometer inside the drum.
The temperature in the box is controlled by a heating resistor regulated by a temperature controller, and is always fixed at 29.0$^{\circ}$C.
Homogeneity of temperature and hygrometry inside the vessel is provided by a fan.
To enable the vapor to penetrate the pile, the drum is filled with the grains, placed in the closed vessel, and rotated at 1 rpm during 12 h.
Just before the experiment, the drum is rotated at 5 rpm during a few seconds to provide an initial reference state.

As the waiting time and waiting angle before the pile destabilization have been found to be crucial parameters, due to capillary bridges nucleation, a fixed waiting time $t_\mathrm{w}$ of 3 s, 1$^\circ$ below the maximum stability angle, has been respected for each experiment \cite{Bocquet,Restagno}.
To reduce the statistical uncertainty, every angle has been measured one hundred times for each of the 12 fine grains contents, 4 relative humidities and 23 rotation speeds.
To check the absence of noticeable segregation, we have sampled the top and bottom of the heap at the end of some experiments, and verified by optical microscopy that they have similar content of fine particles.
To test the presence of electrical charges in the heap, we have measured the voltage of the granular heap surface at a 20 mm distance with a fieldmeter before and just after a rotating drum experiment.
We had not been able to measure an evolution, the voltage keeping a $\sim$50 V value, close to the sensitivity of the device, which can be considered as negligible, usual voltages in charged isolating materials being of the order of kV.

\section{Results}

We have performed rotating drum experiments with a mixture of 250 and 5 $\mu$m glass beads.
By varying the rotation rate $\omega$, the percentage of fine particles $\phi$ (volume ratio of fine and coarse spheres) and the relative humidity $H$, we have built a stability diagram of the heap (see figure \ref{stab_diag} and videos in Supplemental Material \footnote{See Supplemental Material at [URL] for the videos of the avalanching, continuous flow, stick-slip and continuous sliding regimes.}).

\begin{figure}
\centerline{\includegraphics[width=\columnwidth]{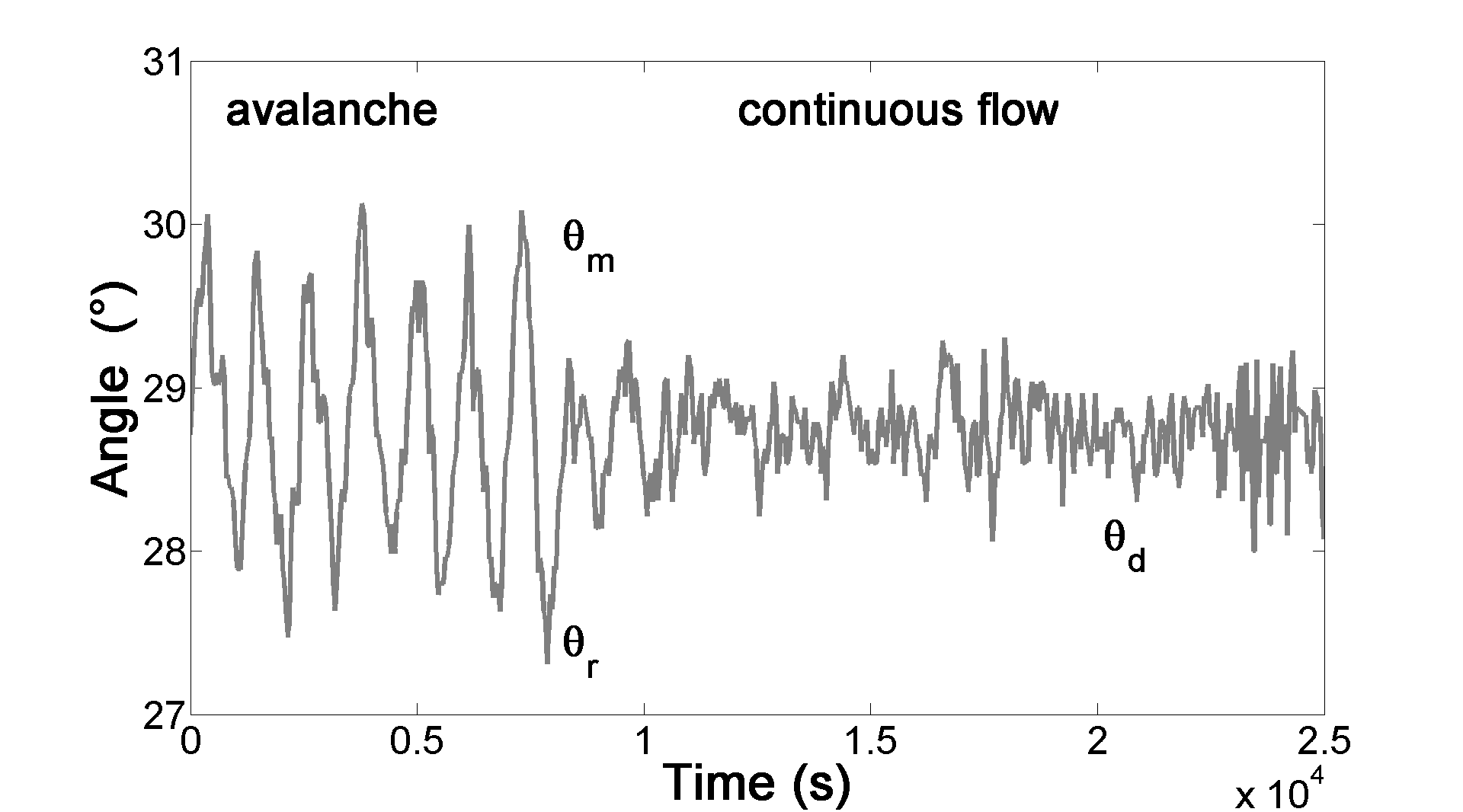}}
\caption{Angle of the surface of the pile as a function of time for 0.002\% of fine particles, 23\% relative humidity and 0.7 rpm rotation rate.
Being in the intermittent regime, avalanching and continuous flow alternate during this experiment.}
\label{angle_time}
\end{figure}

At small concentrations, the rotating heap shows intermittent avalanches, i.e., the top surface remains flat until a maximum stability angle $\theta_\mathrm{m}$ is reached, where the grains flow to achieve a new stable flat surface state at the repose angle $\theta_\mathrm{r}$ (video \#1).
When the rotation speed is increased, the rotation time becomes too short compared to the avalanche duration, and a transition from the avalanche regime to a continuous flow regime is observed, where the grains at the top of the heap flow without rest, keeping a constant flat surface, characterized by the so-called dynamic repose angle $\theta_\mathrm{d}$ (video \#2).
An intermittent regime is observed during the avalanche/continuous flow transition (figure \ref{angle_time}) \cite{Fischer}.
When the fine particles fraction increases, the transition occurs for lower rotation rates (figure \ref{stab_diag}) and the avalanche angle decreases (figure \ref{dra_fines}), proof of a fluidization of the pile, or at least of its surface.

Up to that point, the granular assembly exhibits expected motion patterns \cite{Seiden}.
But at larger fine particles concentration, the pile shows a stabilization of its surface.
It remains now cohesive and, unlike standard rotating drum experiments, the motion of the drum is accommodated by a stick-slip behavior at the drum wall.
When the drum rotates, the pile remains stuck to the wall until a maximum stability angle is attained (as for avalanching), where it detaches and slips downward to the repose angle, without any observable inner motion, except in a layer of grains in contact with the drum wall (video \#3).
When the rotation rate is enhanced, stick-slip stops and a thin layer of grains at the drum wall rolls continuously, thereby leading to an overall sliding motion with a constant slope of the heap surface (video \#4).
In configurations where granular matter exhibits stick-slip at a wall contact, e.g. shearing in a Couette geometry, such a transition between stick-slip and continuous sliding is also observed and has been interpreted as a dilatancy transition \cite{Nasuno}.

\section{Discussion}

So, for low rotation rates, when the fine particles content increases (i) the heap flows at the surface during more and more fluid avalanches, until (ii) this flow is stopped, and the heap slips at the bottom behaving as a quasi-solid body.
This non-monotonous nature of the evolution of the maximum stability angle with fine particles thus enables to bridge a gap between independent observations.
Small particles are known to increase the flowability of powders \cite{Yang} or the run-out length of landslides \cite{Linares-Guerrero}, in weakening the adhesive contact between the coarse ones.
Besides, the introduction in a particulate assembly of particles of smaller diameter is known to promote jamming \cite{Parisi}, although little is known quantitatively about the influence of polydispersity on jamming \cite{Clusel,Pournin}.
This contradictory nature of the influence of a small amount of fine grains on a granular assembly had not been reported yet.
Indeed it had not been studied in one well-defined system, so a coherent view of the influence of dust was not possible.
We mention that a change of behavior is expected when the fine particles begin to dominate the mixture, from $\phi\sim 40\%$ \cite{Naeini,Xenaki}, but was unexpected for values of $\phi$ so low.

\begin{figure}
\centerline{\includegraphics[width=\columnwidth]{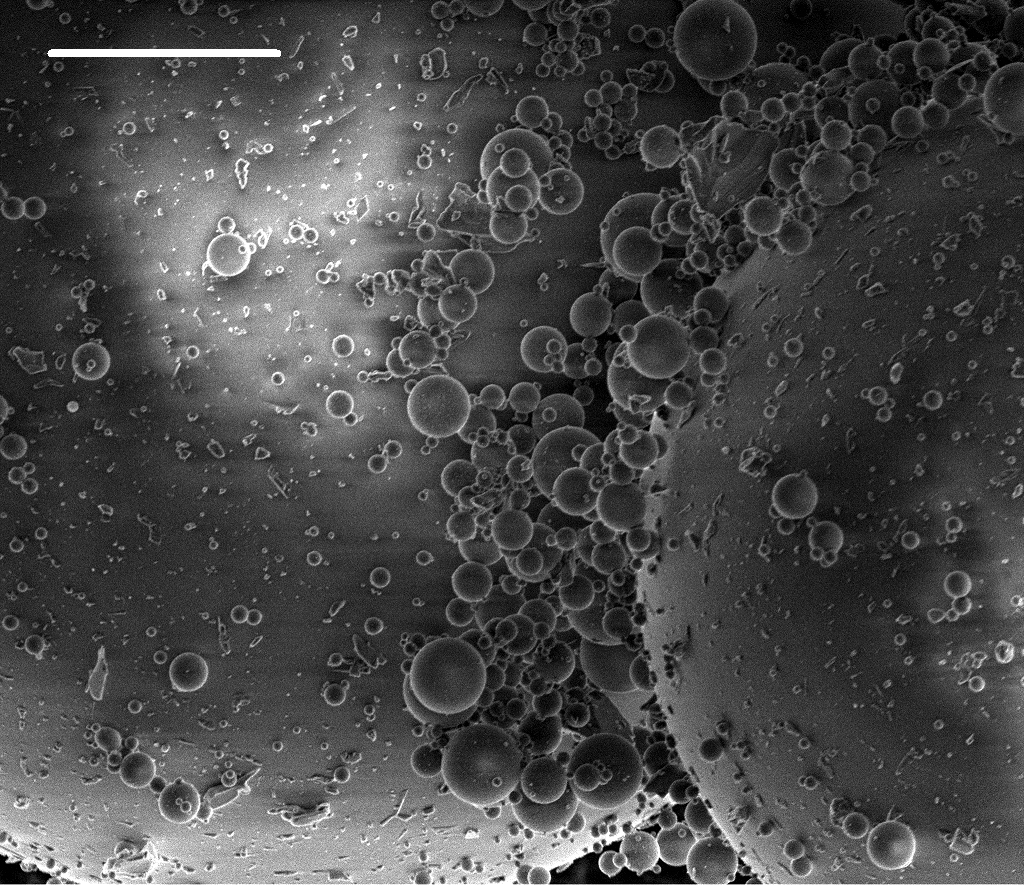}}
\caption{Scanning Electron Microscopy picture of the granular medium after a rotating drum experiment with $\phi=0.15\%$ and $H=43\%$.
The white bar length is 30 $\mu$m.
As performed in vacuum, this observation cannot evidence the capillary bridges.}
\label{MEB}
\end{figure}

\begin{figure}
\centerline{\includegraphics[width=\columnwidth]{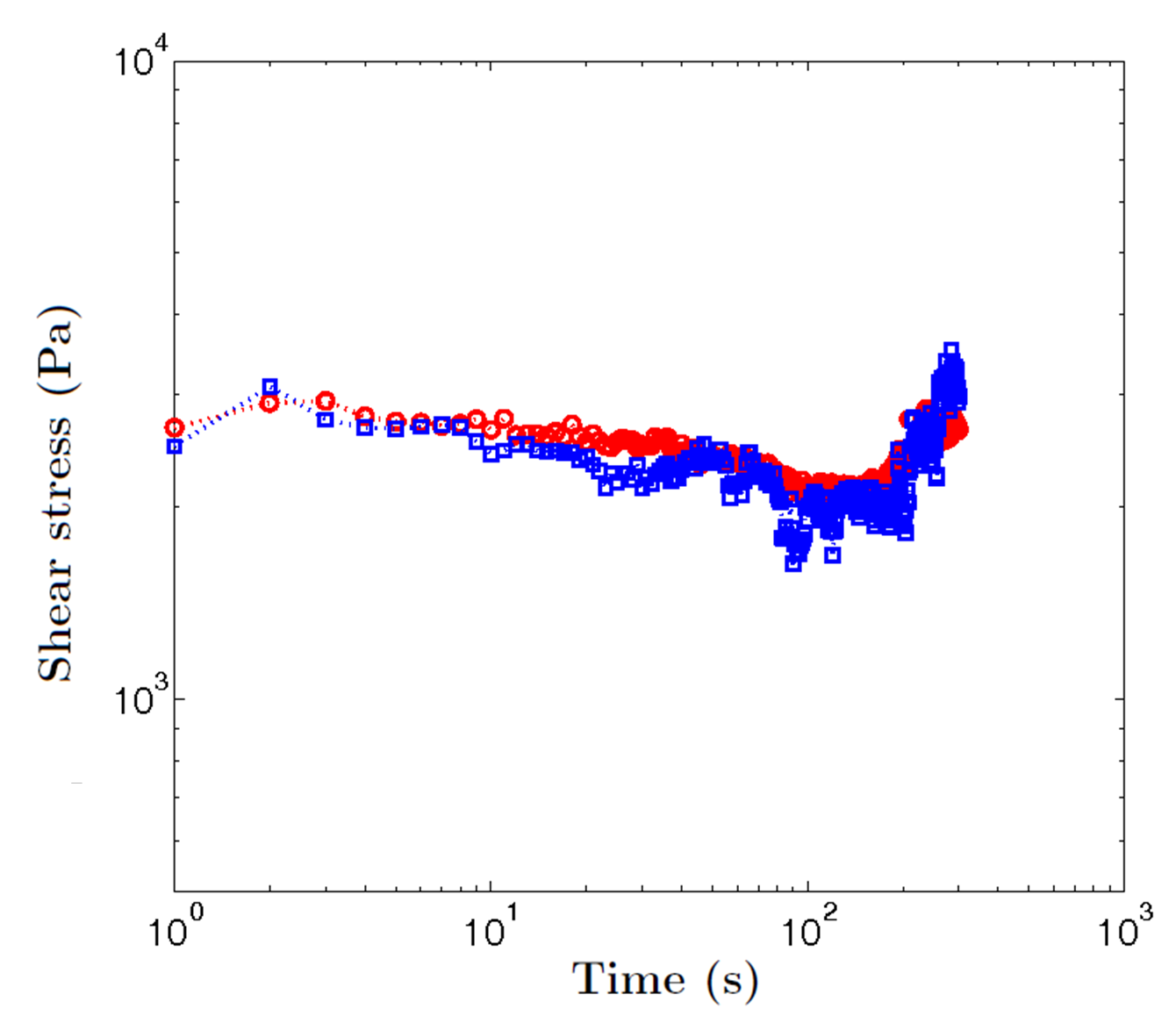}}
\caption{Maximum shear stress measured at the top plate as a function of time, in a parallel plate rheometer containing the humid granular medium and oscillating at 0.1 Hz.
Sandpaper is stuck at both plates to guarantee adhesion of the grains.
The blue squares are the values of the heap without fine grains.
The red squares are the values of the heap with $\phi=1\%$ of fine grains.
We see that the cohesion induced by the fine particles at this concentration does not result in a significant increase of the shear stress needed to strain the heap.}
\label{rheo}
\end{figure}

The locus of destabilization of a granular heap is known to be highly dependent on the cohesion of the particulate assembly, the more cohesive the assembly, the deeper the breakage \cite{Restagno_c}.
When the fine particles volume fraction reaches the critical value $\phi_\text{c}\simeq 0.2\%$, the locus of the breakage moves from the top to the bottom of the heap, proof of a change of cohesion.
The value of $\phi_\text{c}$ is therefore related to the ability of the fine particles to increase the cohesion of the grain assembly.
No prediction of packing geometry in available for our experiments \cite{Parisi}.
Indeed, for such a large bead radius ratio and low fine beads percentage, small particle usually percolate.
So, in order to check the distribution of fine particles inside the network of large beads, we have carried out scanning electron microscopy observations of the mixture just after a rotating drum experiment.
Figure \ref{MEB} clearly indicates that, even if some fine grains are stuck at the surface of big beads, most of them populate the interstitial space between large particles.
The most probable source of cohesion is therefore jamming, promoted by the adhesion of small beads on large ones and between small ones, which prevents them from percolating, and finally blocks the inner motions inside the heap
The influence of this jamming on the mechanical behavior of the heap may be diverse.
Whereas strong in the rotating drum, we have found it much less pronounced in a parallel plate rheometer (figure \ref{rheo}).
In other words, stuck fine particles build kinds of solid bridges, that play the same role as liquid bridges, either fluidifying, or solidifying the heap, as observed by Chou \& Hsiau \cite{Chou}.

\begin{figure}
\centerline{\includegraphics[width=\columnwidth]{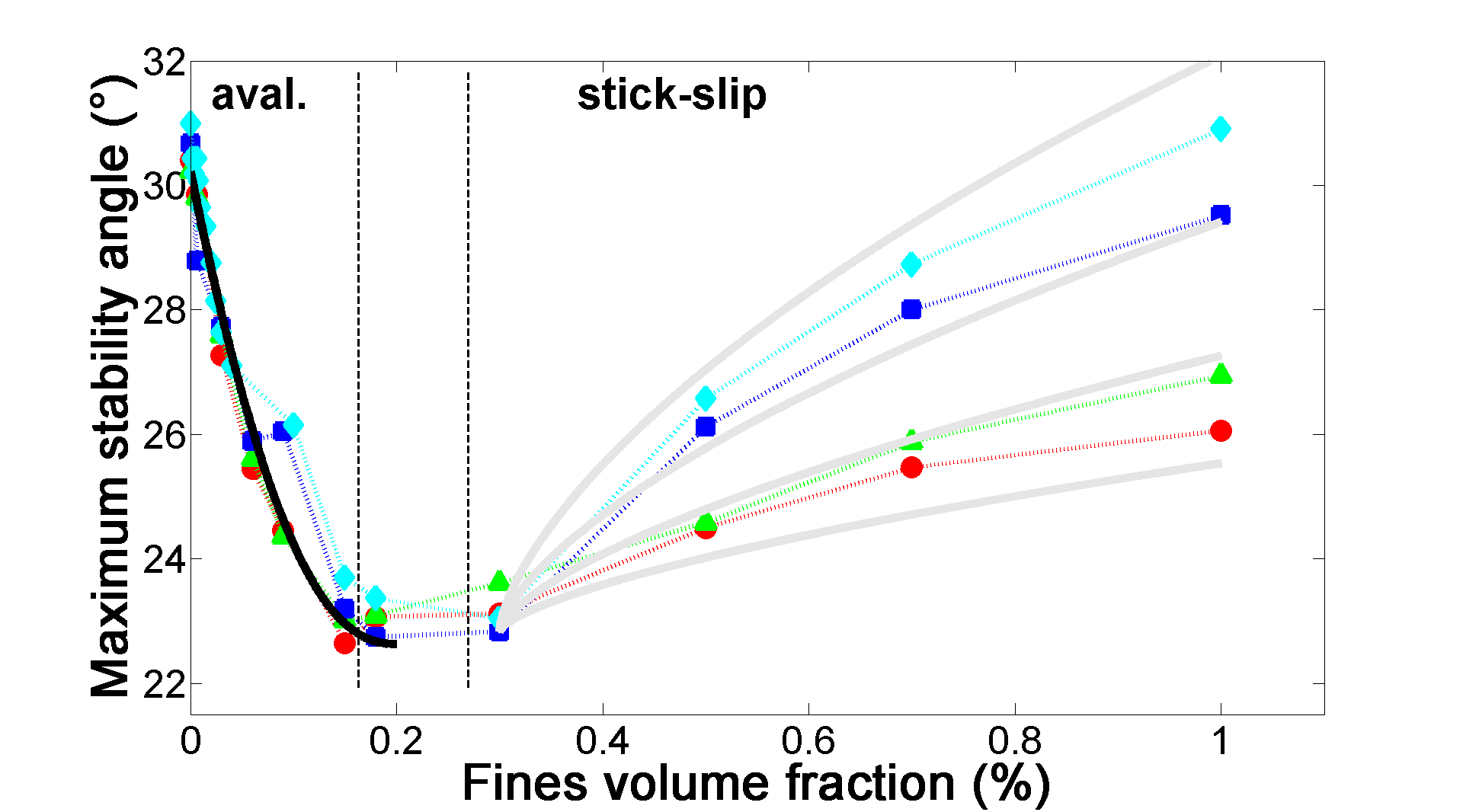}}
\caption{Maximum stability angle of the humid granular heap $\theta_\text{m}$ \textit{vs} percentage of fine particles $\phi$ for $H=6$\% (red circles), 23\% (green triangles), 43\% (blue squares) and 58\% (light blue diamonds) in the avalanche and stick-slip regimes.
In the avalanche regime, the experimental data are fitted by $\tan\theta_\text{m} = \tan\theta_0 (1+A(\phi - \phi_\text{c})^n)$ with $A=13.5$ and $n=2.3$ (black curve).
In the sliding regime, all the experimental data are fitted by Equation \ref{slideq} with $B=3.3$ (grey curves).}
\label{dra_fines}
\end{figure}

For low fine grains content, the avalanche regime is characterized by three features: (i) the destabilization always occurs at the surface of the heap, (ii) the maximum stability angle is influenced by the fine particles content (so the cohesion) but (iii) not by the relative humidity (figure \ref{dra_fines}).
This behavior is typical of the so-called granular regime \cite{Tegzes}.
In this regime, as the real contact surface between beads (so the adherence force between them) increases proportionally with the depth inside the heap, due to elastoplatic strains, the most unstable plane remains at the surface \cite{Restagno_c}.
And for our relative humidity values, the equilibrium radius of curvature of the water menisci condensing between beads is smaller than their typical roughness ($\sim 10\,\text{nm}$).
So the capillary water is confined to the tip of asperities in contact between beads and the adhesion force between them is not influenced by humidity, as theoretically described \cite{Bocquet02} and experimentally observed \cite{Fraysse} in monodisperse media.
In this regime, the maximum stability angle obeys the simple law: $\tan(\theta_\text{m})=\tan\theta_0(1+C)$, with $\theta_0$ the angle in absence of cohesion and $C$ a function of the adhesion stress inside the heap \cite{Restagno_c}.
The theoretical value of $\theta_0$, computed from pure geometrical considerations, is 23.4$^\circ$ in a monodisperse system \cite{Albert}.
We see in figure \ref{dra_fines} that the minimum experimental value of $\theta_\text{m}$, measured at $\phi_\text{c}$, corresponds to this value.
We can deduce from this equality that, before jamming the heap at high content, the fine spheres begin by playing the role of ball bearings.
By populating the intergrain space (figure \ref{MEB}), they contribute to separate the beads and destroy their cohesion, inducing granular lubrication, until recovering the behavior of a cohesion-free heap, when $\theta_\text{m}$ reaches $\theta_0$.
By fitting the experimental results with the above law, we can conjecture that the loss of cohesion, so the granular lubrication efficiency, due to the fine particles, follows a scaling law of the form $C \sim (\phi - \phi_\text{c})^n$ with $n\simeq 2.3$ (figure \ref{dra_fines}).

For high fine particles contents, the granular assembly now behaves as a solid body and its sliding at the drum wall is driven by the friction between the glass spheres and the steel cylinder.
The maximum stability angle $\theta_\text{m}$ can thus be deduced from a standard Coulomb analysis.
At first order, we perform this analysis at the location of the drum wall with the tangent parallel with the pile surface.
Here the humid granular heap loses its stability when the ratio of the shear stress $P\sin\theta_\text{m}$ ($P$ the weight of the above grains) to the normal stress $P\cos\theta_\text{m}+c$ ($c$ the pile-wall adhesion force per unit surface) exceeds the friction coefficient $\mu=\tan\theta_\text{s}$ \cite{Halsey}.
As $P$ is $\rho g h$, with $\rho$ the density of the heap, $g$ the acceleration of gravity and $h$ the height of the heap above the considered point, the stability criterion can therefore be written: $\tan\theta_\text{m} = \tan\theta_\text{s}(1+c/(\rho g h \cos\theta_\text{m})).$
The increase of adhesion with $\phi$ results from the capillary bridges nucleating between the fine grains and the wall, bonding them to the drum.
The enhancement of the density of fine particles increases their number at the drum surface, and thereby increases the total number of capillary bridges and the pile-drum adhesion.
If we consider that at the minimum value of the sliding angle, for $\phi_\text{m}\simeq 0.3\%$, we are in the adhesion-free case, then at this point $c=0$ and $\theta_\text{m}=\theta_\text{s}$.
The adhesion stress $c$ is linked to the number of fine particles per unit surface, so it can be written $c=A(\phi-\phi_\text{m})^{2/3}$, with $A$ a factor linked to the beads size and roughness and to the relative humidity $H$.
The dependence of $A$ on $H$ is $A=B/\ln(1/H)$, $B$ being now a purely geometric factor \cite{Bocquet}.
Therefore the stability of the pile writes:
\begin{equation}
\tan\theta_\text{m} = \tan\theta_\text{s}\left(1+\frac{B}{\rho g h \cos\theta_\text{s} \ln(1/H)}(\phi - \phi_\text{m})^{2/3}\right)
\label{slideq}
\end{equation}
Here, following Halsey and Levine, we have considered that $\cos\theta_\text{m}\simeq\cos\theta_\text{s}$ to get a more tractable expression \cite{Halsey}.
The fitting in figure \ref{dra_fines} of all the experimental sliding angles with this law, using $B$ as only fitting parameter, enables to catch satisfactorily the influence of humidity and the curved shape of the angle evolution.
Thus considering that the enhancement of friction between the particulate assembly and the wall originates from the nucleation of capillary bridges gives the correct scaling for the influence of the fine particles content and humidity, thereby validating the assumption.

\section{Conclusion}

We have shown here that the presence of dust in a monodisperse granular assembly modifies its cohesion for a content as low as a fraction of percent.
Furthermore this influence is ambivalent, the fine grains firstly lubricating the pile, and then jamming it.
This modification of the inner dynamics of the pile has a huge impact on its mechanical stability.
In our rotating drum configuration, it induces a change of destabilization mode, from avalanching to stick-slip.
It may induce different changes in other configurations.
This study helps to reconcile independent and contradictory knowledges about the fine particles influence.
The implications of this decisive role of dust are direct.
In industrial processes, the fine grains produced at the beginning of mineral grinding for instance appear here to be likely to lower the process efficiency.
Beside, 
this system, despite simple, has shown to be an original way of investigating the nature of arrested states (what is the exact mechanism of jamming in a strongly bidisperse mixture?), the interaction inside polydisperse aggregates (what drives the granular lubrication efficiency?), or the tribology of humid granular matter (how does a bidisperse pile interact with the container walls?).

\vspace{0.5cm}

\section*{Acknowledgments}

We thank Elisabeth Charlaix, Lyd\'eric Bocquet and Laurent Joly for enlightening discussions, Catherine Barentin and Loren Jorgensen for the rheometric measurements, Fran\c{c}ois Gay and Matthieu Guibert for experimental help, and the Carnot Institute Ing\'enierie@Lyon for financial support.

\begin{figure}
\centerline{\includegraphics[width=\columnwidth]{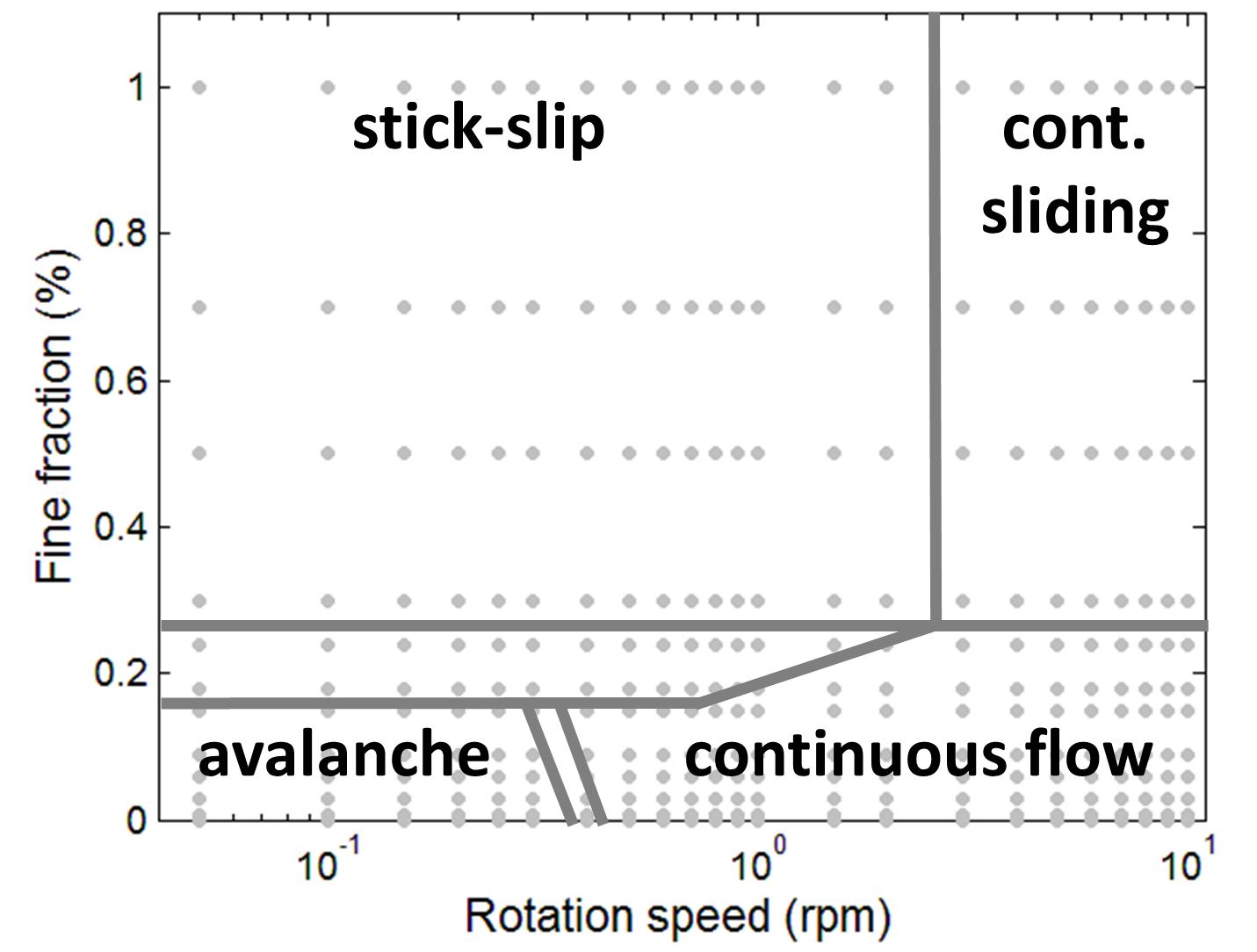}}
\caption{Stability diagram of the granular heap.
All regimes are shown as a function of the rotation velocity of the drum and of the percentage of fine particles for a relative humidity $H = 23\%$.
The well-defined regimes are separated by transition domains showing intermittent regimes.
Each of the 276 dots corresponds to an average of 100 measurements.
The increase of humidity induces exclusively a slight shift of the avalanche/continuous flow and stick-slip/continuous sliding transitions toward the high rotation rate.}
\label{stab_diag}
\end{figure}

\end{document}